\pdfoutput=1
\documentclass[sigconf]{acmart}

\AtBeginDocument{%
  }

\copyrightyear{2026}
\acmYear{2026}
\setcopyright{cc}
\setcctype{by}
\acmConference[SA Technical Communications '26]{SIGGRAPH Asia 2026 Technical Communications}{December 01--04, 2026}{Kuala Lumpur, Malaysia}
\acmBooktitle{SIGGRAPH Asia 2026 Technical Communications (SA Technical Communications '26), December 01--04, 2026, Kuala Lumpur, Malaysia}
\acmDOI{10.1145/3829339.3847832}
\acmISBN{979-8-4007-2841-9/2026/12}

\usepackage{subcaption}

\begin{document}

\title{Sensitivity as an Arbitrary Output Variable for Differentiable Rendering}

\author{Linas Beresna}
\orcid{0009-0004-7102-324X}
\affiliation{%
  \institution{Simon Fraser University}
  \city{Burnaby}
  \state{BC}
  \country{Canada}}
\email{linas\_beresna@sfu.ca}

\author{Eugene Fiume}
\orcid{0000-0001-5939-7055}
\affiliation{%
  \institution{Simon Fraser University}
  \city{Burnaby}
  \state{BC}
  \country{Canada}}
\email{eugene\_fiume@sfu.ca}

\renewcommand{\shortauthors}{Beresna and Fiume}

\begin{abstract}
Differentiable renderers expose the derivative of any scalar objective with
respect to every scene parameter, yet unlike the primal image, which decades of
arbitrary output variables (AOVs) have taught us to decompose, inspect, and
composite, these derivatives have no established representation for human
inspection. We introduce the \emph{sensitivity AOV}, a render output carrying
the sensitivity of an objective to the scene parameters that influence it. A
single reverse-mode pass populates a \emph{sensitivity buffer} over the scene's
parameter hierarchy, from which many views are read rather than
re-differentiated, in direct analogy to deferred shading: image-space
sensitivity at object and parameter-type granularity, projections onto a freely
navigable scene from any inspection viewpoint, and, for spatially varying
parameters, per-texel fields carried to the surface through texture
coordinates. We separate the fixed camera that defines the objective from the
free camera used to inspect the result, and position reverse-mode attribution
against its forward-mode dual. Our aim is not a single algorithm but a
scaffolding that establishes derivative outputs as first-class render products
alongside the primal image.
\end{abstract}




\begin{teaserfigure}
  \centering
  \includegraphics[width=\linewidth]{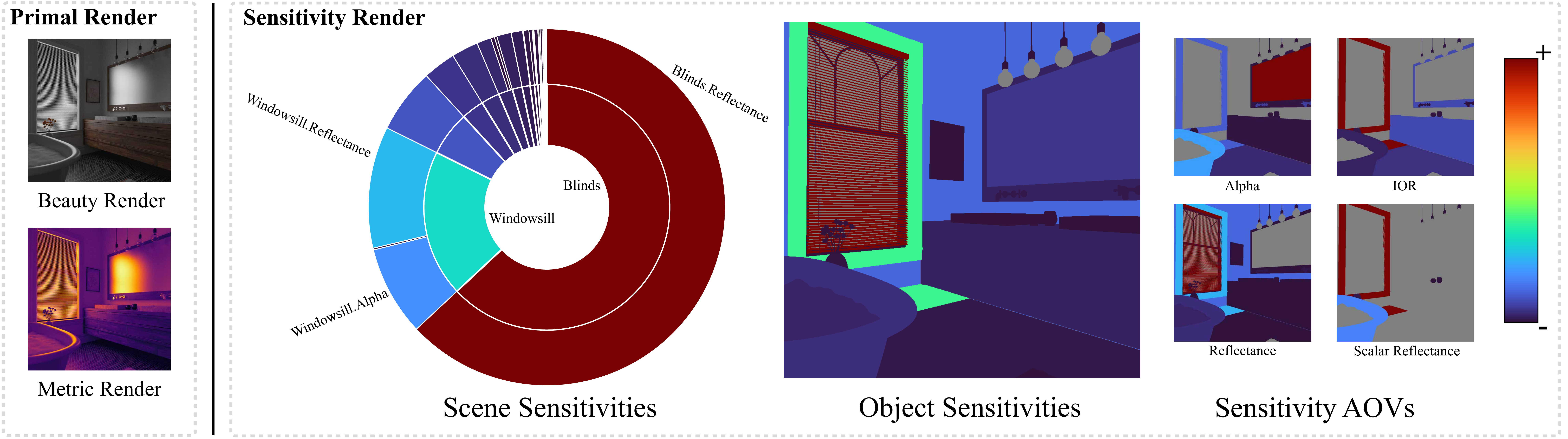}
  \caption{Sensitivity AOVs. Beside the primal outputs (left), one reverse-mode
pass yields the derivative of a scalar objective with respect to every scene
parameter (right). One buffer supports several views: a sunburst ranking the
most influential parameters, a projection onto primary visibility locating them
in the image, and per-type channels, each signed, blue where increasing the
parameter lowers the objective and red where it raises it.}
  \Description{A bedroom interior scene shown in two groups. On the left, the
primal outputs: a beauty render and a metric render of the same room. On the
right, three views of one sensitivity buffer. The first is a two-ring sunburst
chart whose inner ring names objects, Blinds and Windowsill, and whose outer
ring splits them into parameters such as Blinds reflectance, Windowsill
reflectance and Windowsill alpha, with Blinds reflectance taking the largest
share. The second is the room re-coloured per object, with the window blinds
and a floor region highlighted in bright green against a blue background. The
third is a grid of four per-type channels labelled Alpha, IOR, Reflectance and
Scalar Reflectance, each drawn with a signed blue-to-red colour map whose bar
runs from minus to plus.}
  \label{fig:teaser}
\end{teaserfigure}

\maketitle

\begin{figure*}[t]
\centering
  \includegraphics[width=\linewidth]{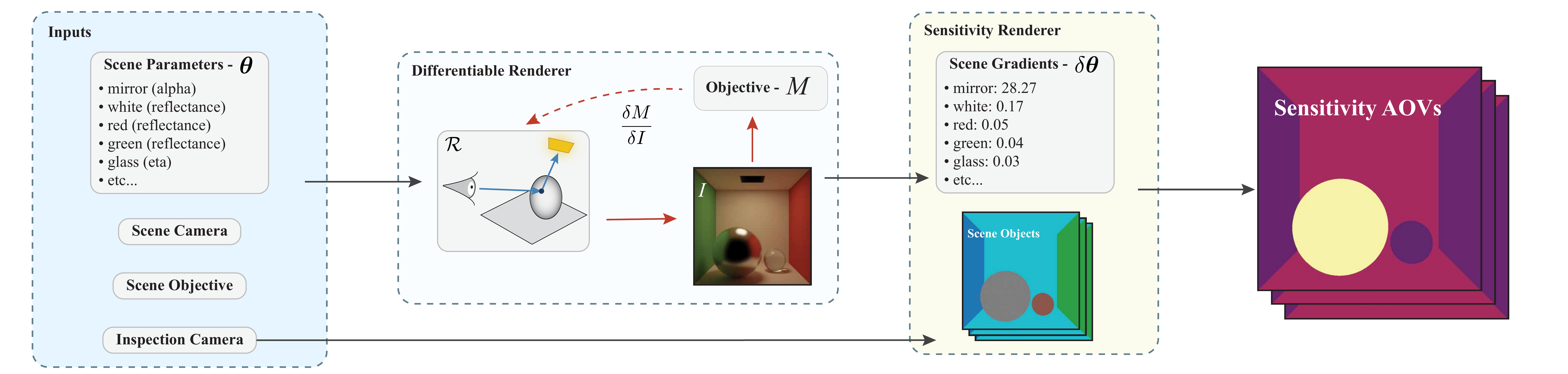}
\caption{The sensitivity AOV pipeline. The inputs (left) are the scene
parameters $\theta$, the camera and objective defining the differentiated
quantity, and an \emph{inspection camera} used only for viewing. The renderer
(centre) forms $I=\mathcal{R}(\theta)$ and evaluates $M$; one reverse-mode pass
propagates $\partial M/\partial I$ back through light transport to yield a
sensitivity for every parameter (right). That populated set is the sensitivity
buffer: binding it to primary visibility gives the \emph{Scene Objects} pass,
selecting subsets over it the \emph{Sensitivity AOVs}.}
\Description{Pipeline diagram from objective camera through the adjoint pass
to the sensitivity buffer and the family of views.}
\label{fig:architecture}
\end{figure*}

\section{Introduction}
\label{sec:introduction}

Production renderers rarely output only the final picture. Alongside the
beauty pass, an \emph{arbitrary output variable} (AOV) exposes some
intermediate quantity of the light transport, such as depth, surface normals,
albedo, or object and material identity, so that artists can inspect, relight,
denoise, and composite~\cite{openexr}. Decades of practice have made these
outputs first-class: a compositor expects a normal pass, and an identity pass
such as cryptomatte~\cite{cryptomatte2015} is a standard deliverable.

Differentiable renderers~\cite{LiEdge2018, NimierDavidVicini2019Mitsuba2,
NimierDavid2020Radiative, Vicini2021PathReplay} expose a fundamentally new
quantity: the derivative of any scalar objective evaluated on the image with
respect to \emph{every} scene parameter. A single reverse-mode pass returns,
for each material, light, and geometric parameter, how sensitive the objective
is to that parameter. To date this quantity has been consumed almost
exclusively by optimizers, in inverse rendering and design. Yet the derivative
is also informative on its own: it tells a human which parts of a scene matter
for a given goal. Previous work~\cite{beresna2026saliency} established that
these gradients are interpretable and that the resulting rankings are
objective-specific, but treated each map as a one-off visualization; what the
quantity still lacks, and what is supplied here, is the output representation
the primal image has long possessed, a reusable buffer and a taxonomy of the
views read from it.

Treating the gradient of an output as an attribution over its inputs is by now
standard in neural network interpretability~\cite{selvaraju2017grad}. We apply
the principle to a different differentiable function: the inputs become scene
parameters and the output an arbitrary scalar objective, so the attribution
lands in the scene rather than in a feature map.

We propose the \emph{sensitivity AOV}: a render output carrying the sensitivity
of an objective to the scene parameters that influence it. Where a forward AOV
is \emph{scene-intrinsic} (depth is depth, independent of any goal), a
sensitivity AOV is \emph{objective-conditioned}, so the same scene yields
different sensitivity AOVs under different objectives. Forward AOVs answer
\emph{what is here}; sensitivity AOVs answer \emph{what would change this}.

Our contribution is a scaffolding rather than a single algorithm. We: (i) define
the sensitivity AOV along four axes: selection, binding, aggregation, and
encoding; (ii) observe that one reverse-mode pass populates a \emph{sensitivity
buffer} over the scene's parameter hierarchy, from which many views follow
without recomputation, in direct analogy to deferred shading; (iii) give a
taxonomy of those views, distinguishing where a parameter \emph{lives} from
where its influence \emph{lands}; and (iv) provide a reference implementation.
Our aim is to establish derivative outputs as first-class render products
alongside the primal image.

\section{The Sensitivity Buffer}
\label{sec:buffer}

Let $\mathcal{R}$ be a differentiable renderer producing an image
$I=\mathcal{R}(\boldsymbol{\theta})$ from scene parameters
$\boldsymbol{\theta}$, and let $M$ be a scalar objective evaluated on $I$. A
single reverse-mode pass yields
\begin{equation}
\mathbf{g}=\frac{\partial M(\mathcal{R}(\boldsymbol{\theta}))}
{\partial\boldsymbol{\theta}}
=\frac{\partial M}{\partial I}\,\frac{\partial I}{\partial\boldsymbol{\theta}},
\label{eq:chain}
\end{equation}
a vector assigning to every scene parameter a scalar sensitivity. Crucially,
$\mathbf{g}$ is computed \emph{once}, at the cost of one backward pass whose
cost is largely independent of the number of parameters. We call the populated
parameter set the \emph{sensitivity buffer}. Every visualization we describe is
a re-projection of this buffer, never a recomputation, the same relationship
deferred shading has to the G-buffer~\cite{Takafumi1990G}: the adjoint sweep
happens once, and the user then explores many views over the stored result.

Scene parameters are naturally \emph{hierarchical}. Renderers expose them
through structured keys such as \texttt{floor.bsdf.reflectance}, so the
buffer is a tree: leaves are individual parameters, interior nodes are
materials and objects.

\section{From Buffer to Image}
\label{sec:aov}

The buffer of Section~\ref{sec:buffer} is a tree of scalar sensitivities, one
per parameter component. A \emph{sensitivity AOV} is what results from reading
that tree at a chosen granularity and drawing it: we select a subtree,
aggregate its components to a single scalar per group, bind that scalar to
pixels, and encode it for display. The first two steps decide \emph{what} is
shown, the last two \emph{where} and \emph{how}.

\begin{figure}[t]
\centering
\includegraphics[width=\columnwidth]{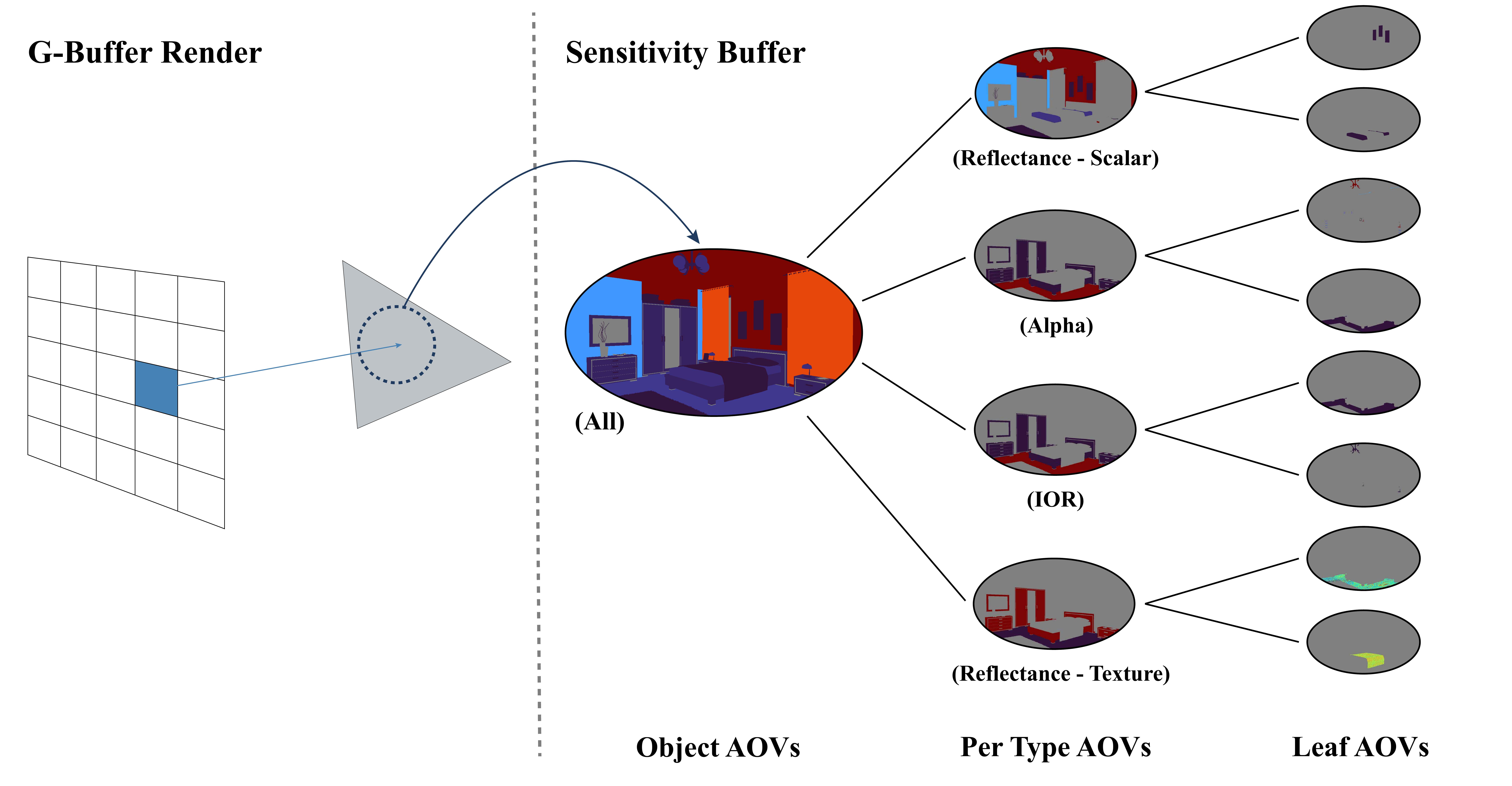}
\caption{The sensitivity buffer as a hierarchy. Left: a primary ray from a
pixel identifies the visible surface, the G-buffer step that binds
sensitivities to the image. Right: the buffer read at three granularities, leaf
parameters aggregating into per-type channels, which aggregate into the
object-level view. Each level is a coarser query over one buffer.}
\Description{A two-part diagram. On the left, labelled G-Buffer Render, a ray
leaves one highlighted pixel of an image grid, passes through a triangle
representing the visible surface, and arrows across to the right-hand part. On
the right, labelled Sensitivity Buffer, a tree fans out through three levels: a
single Object AOV node showing the whole room, branching into four Per Type AOV
nodes labelled Reflectance-Scalar, Alpha, IOR and Reflectance-Texture, each of
which branches again into two Leaf AOV nodes holding sparse, mostly empty
images. Every node is the same scene drawn at a coarser or finer granularity.}
\label{fig:buffer}
\end{figure}

\paragraph{Selection and aggregation.} Selection is a query over the tree, and
two cuts are natural. A \emph{vertical} cut groups by object, collecting every
parameter of one material into a single value, and answers which material
matters (the \emph{Object Sensitivities} of Fig.~\ref{fig:teaser}). A
\emph{horizontal} cut groups by parameter type across the whole scene, yielding
the reflectance, roughness, and index-of-refraction channels shown as the
\emph{Sensitivity AOVs}. Either way a group must reduce to one scalar. We use
the $\ell_2$ norm of the group's value-weighted gradients, which measures the
total response to perturbing the whole group. Because parameters differ in
dimensionality, a single reflectance versus a million-texel texture, this total
can instead be taken per component when the intent is to compare across
parameter types rather than total leverage; the choice is part of the encoding
and is recorded with the channel (Fig.~\ref{fig:buffer}).

\begin{figure}[t]
\centering
\begin{subfigure}{0.32\columnwidth}
  \includegraphics[width=\linewidth]{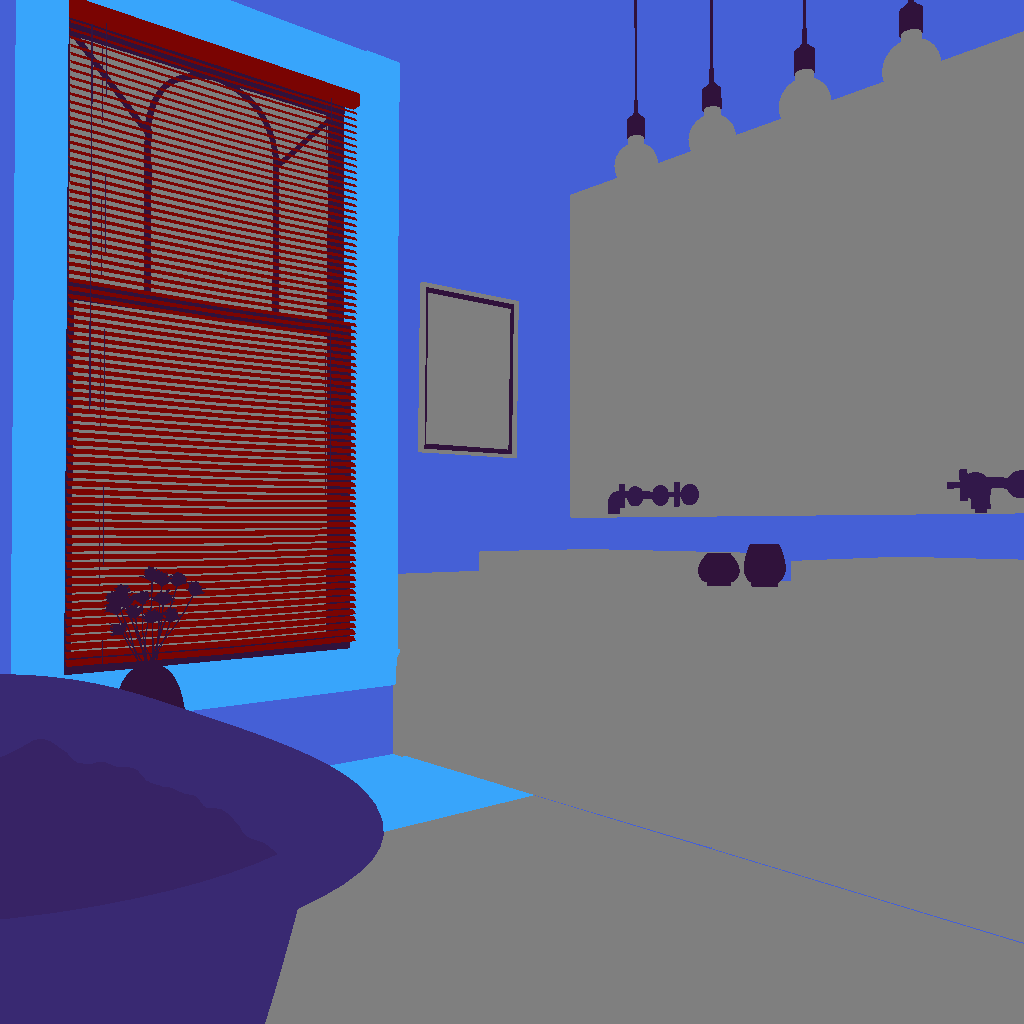}
  \caption{Per object}
  \label{fig:texture-scalar}
\end{subfigure}
\hfill
\begin{subfigure}{0.32\columnwidth}
  \includegraphics[width=\linewidth]{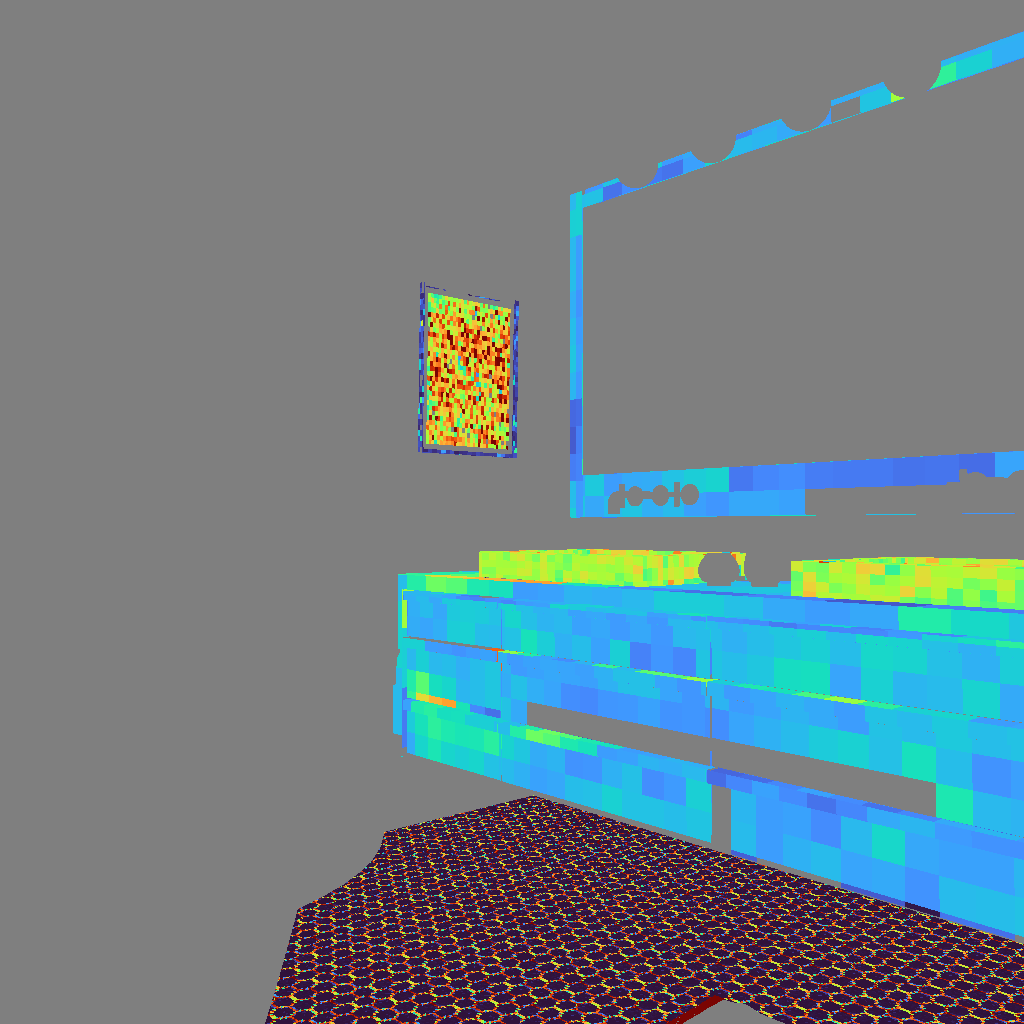}
  \caption{Per texel}
  \label{fig:texture-texel}
\end{subfigure}
\hfill
\begin{subfigure}{0.32\columnwidth}
  \includegraphics[width=\linewidth]{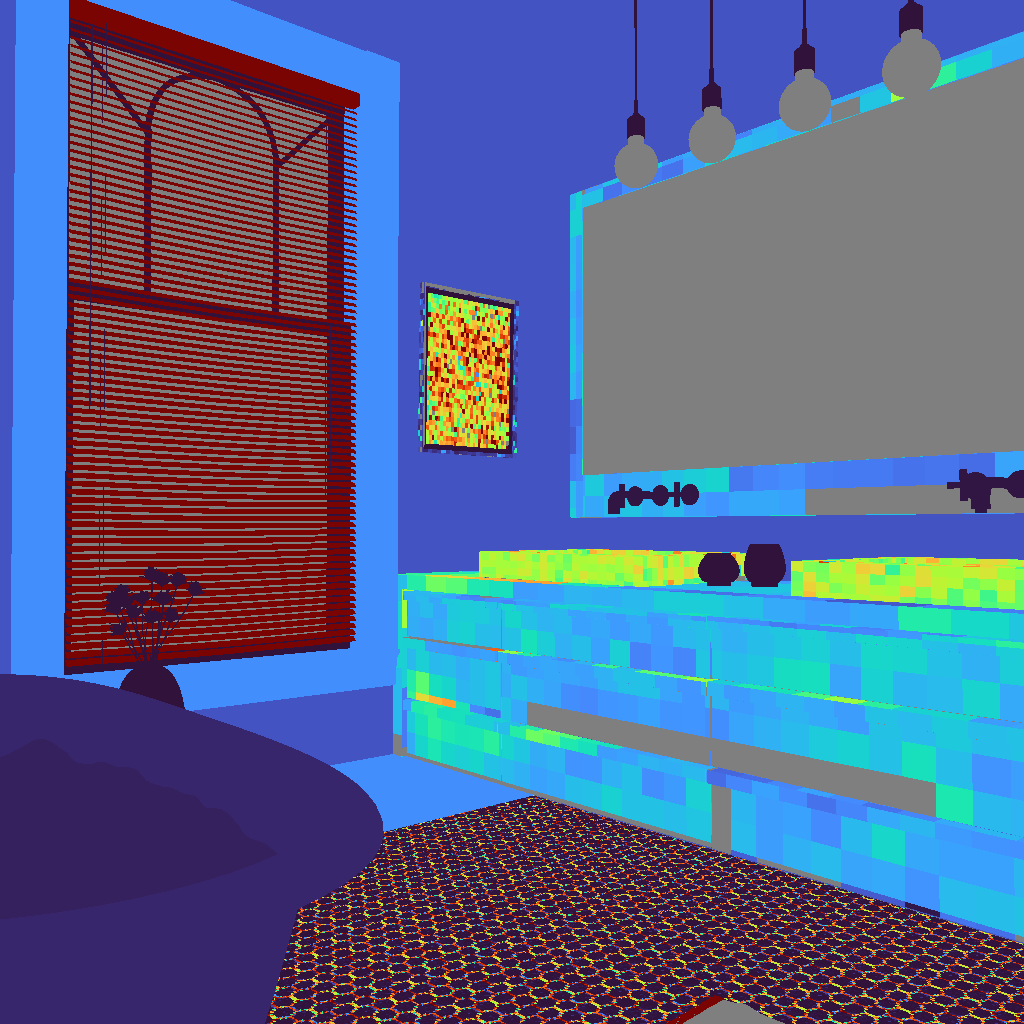}
  \caption{Combined}
  \label{fig:texture-combined}
\end{subfigure}
\caption{Viewing a spatially-varying parameter. A textured reflectance
aggregated to one object value, painted per texel, and under a joint
normalization. The two levels use separate scales (Section~\ref{sec:aov}).}
\Description{Three panels of a textured reflectance: object-level aggregate,
per-texel field, and a combined normalisation.}
\label{fig:texture}
\end{figure}

\paragraph{Binding.} An aggregated value acquires a pixel location through
primary visibility: we trace one ray per pixel, identify the object hit, and
paint that object's sensitivity. Because this step touches only primary rays and
reuses the stored buffer, it carries no gradient computation, so the same
sensitivities can be re-bound from any \emph{inspection camera} without a new
adjoint pass. The objective is fixed by one camera; the camera used to look at
the result is free, and moving it does not change the objective
(Fig.~\ref{fig:multiview}). A spatially varying parameter binds differently: a
texture has its own domain, so its per-texel gradient can be carried to the
surface through the hit UVs and shown in place. This exposes a choice a flat
parameter never forces, since the same texture can be reduced to one object
value or painted per texel, and the two do not share a scale, a single texel's
influence being small beside a whole material's. We treat them as distinct
views of one buffer (Fig.~\ref{fig:texture}) rather than forcing them into a
single image.

\paragraph{Encoding.} A sensitivity can be shown as a signed quantity or a
magnitude, under linear or compressive tone mapping, against a chosen
normalization. The most consequential choice is what a unit of the parameter
means: a raw gradient $\partial M/\partial\theta$ is expressed per unit of
$\theta$, but a unit differs for a reflectance in $[0,1]$ and an unbounded
emitter radiance, so weighting by the value, $\theta\,\partial M/\partial\theta$,
instead reports the response to a proportional perturbation. Because a
sensitivity AOV is objective-conditioned, two maps are comparable only if they
share this encoding, so we record it explicitly with each channel.

\section{Forward and Reverse}
\label{sec:duality}

The buffer is a reverse-mode object: one objective propagated to every
parameter at once, dense in parameters and sparse in objectives. Its transpose
is forward mode, which pushes one chosen parameter through to every pixel,
$\partial I/\partial\theta_i$. The two answer different questions, reverse for
attribution (which parameters matter for this objective) and forward for
prediction (what a single parameter does to the image), and the second cannot
be assembled from the first without repeating the computation per parameter.
Figure~\ref{fig:duality} places both side by side: forward mode is not an
alternative to the buffer but its dual, suited to the question the buffer does
not answer.

\section{Example Objective and Results}
\label{sec:results}
To keep the objective an \emph{example} rather than a contribution, we use the
simplest scalar on the image, its mean photometric response,
$M=\tfrac{1}{|P|}\sum_{p\in P} Y(I_p)$, where $Y(I_p)$ is the photometric
response of pixel $p$. We choose it because its sensitivities are easy to reason
about, which makes the buffer and its views straightforward to check; the
pipeline is agnostic to $M$, and a perceptual or learned objective enters at
exactly the same point, changing only $\partial M/\partial I$.

\begin{figure}[t]
\centering
\begin{subfigure}{0.32\columnwidth}
  \includegraphics[width=\linewidth]{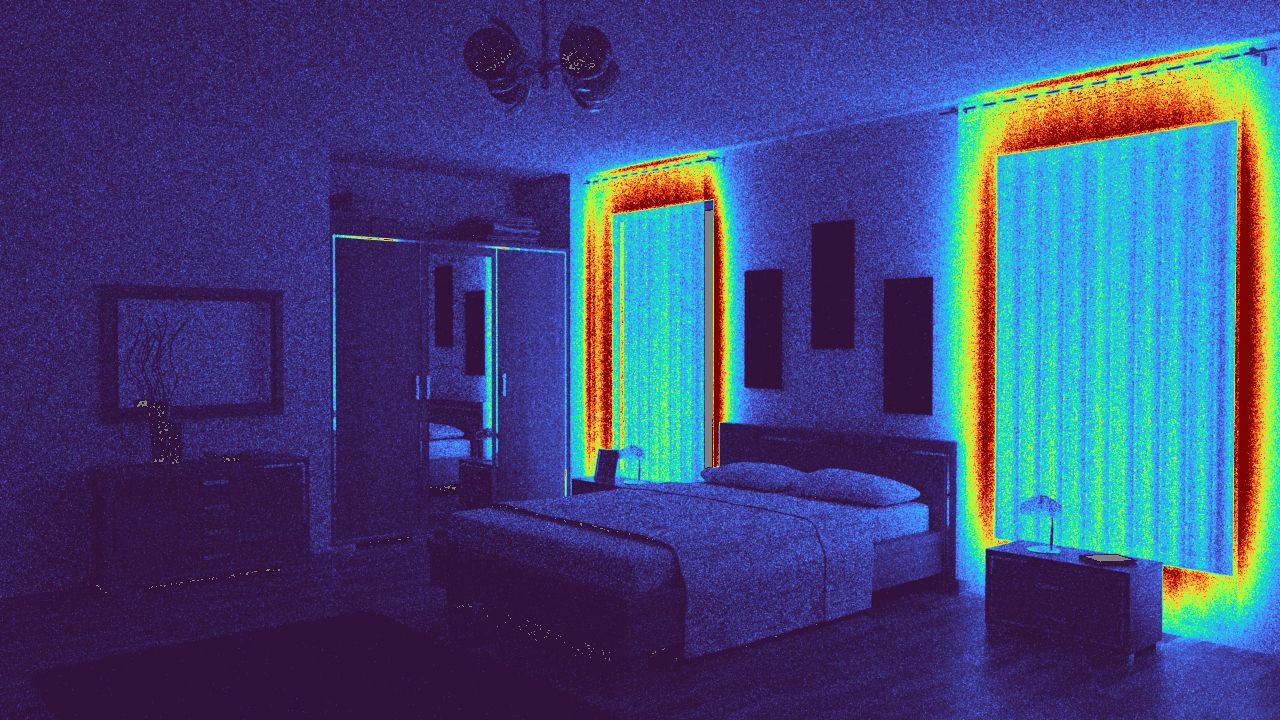}
  \caption{frwd | reflectance}
  \label{fig:duality-fwd-refl}
\end{subfigure}
\hfill
\begin{subfigure}{0.32\columnwidth}
  \includegraphics[width=\linewidth]{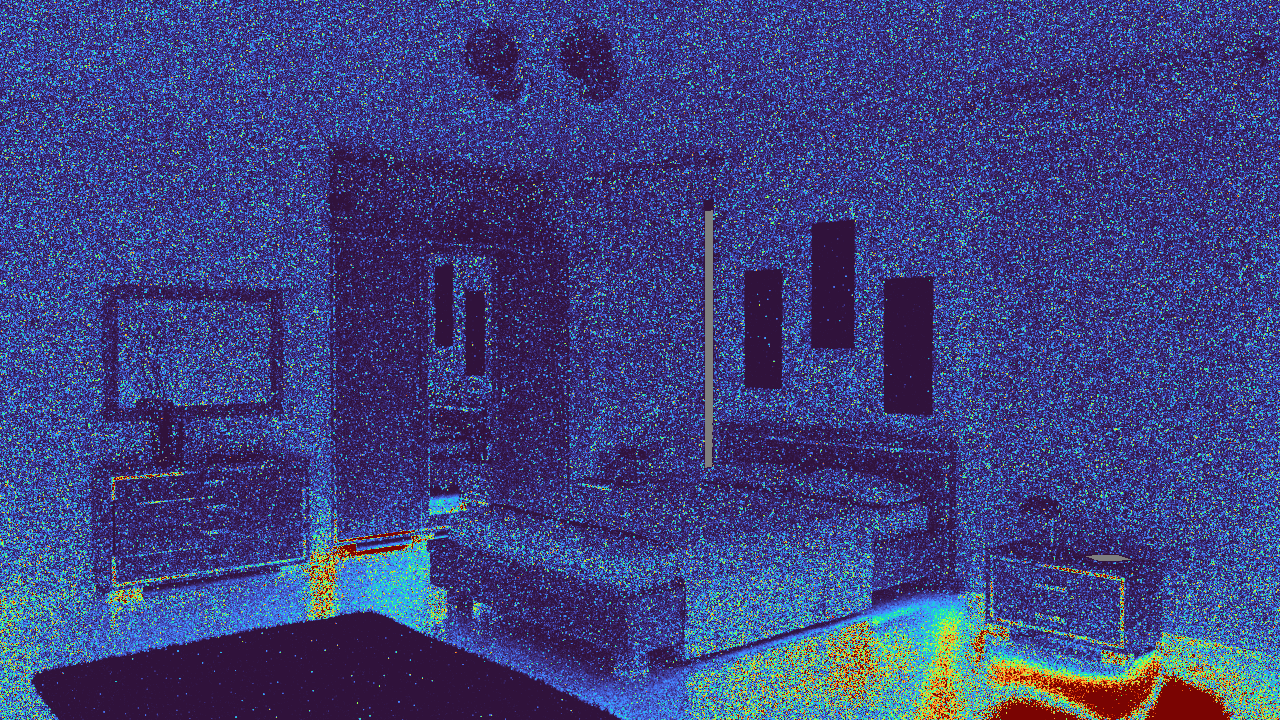}
  \caption{frwd | alpha}
  \label{fig:duality-fwd-alpha}
\end{subfigure}
\hfill
\begin{subfigure}{0.32\columnwidth}
  \includegraphics[width=\linewidth]{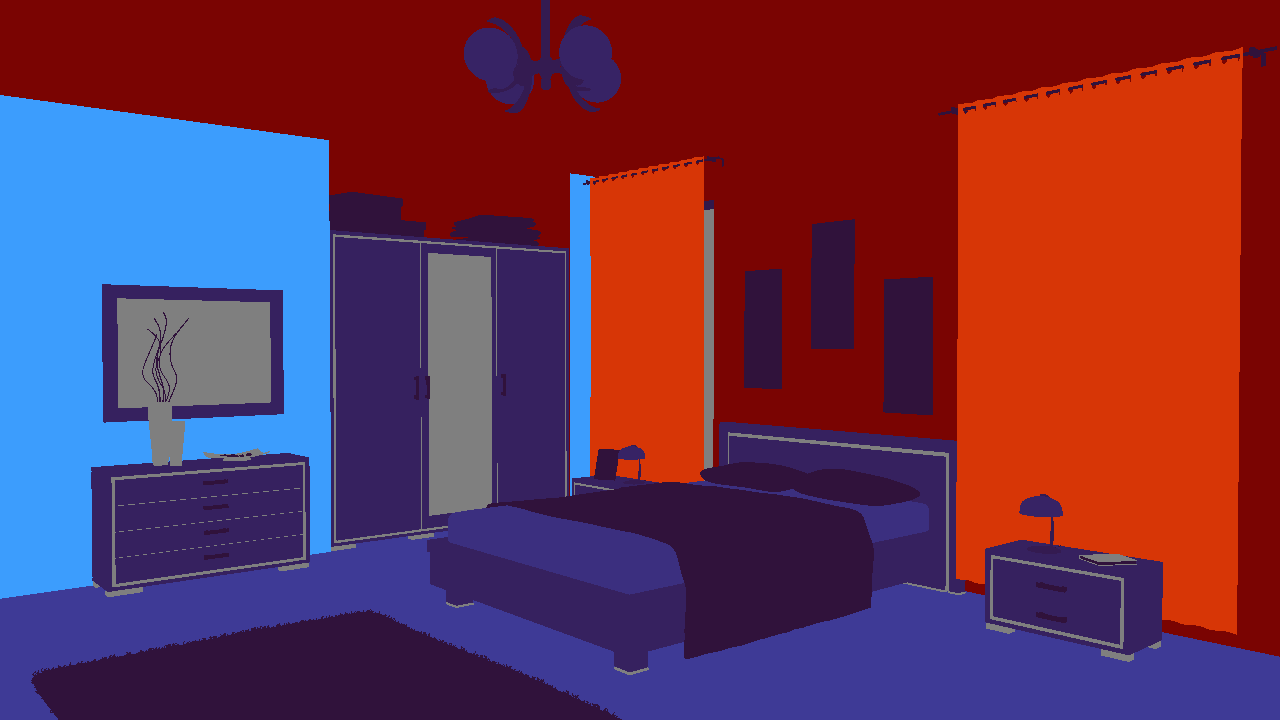}
  \caption{Reverse}
  \label{fig:duality-rev}
\end{subfigure}
\caption{Reverse and forward modes are duals. Forward mode
(a,~b) pushes a single
parameter to all pixels; reverse mode (c) gives one
objective's sensitivity to all parameters at once (Section~\ref{sec:duality}).}
\Description{Three panels: two forward-mode derivative images for single
parameters and one reverse-mode object-level sensitivity map.}
\label{fig:duality}
\end{figure}

\begin{figure}[t]
\centering
\begin{subfigure}{0.32\columnwidth}
  \includegraphics[width=\linewidth]{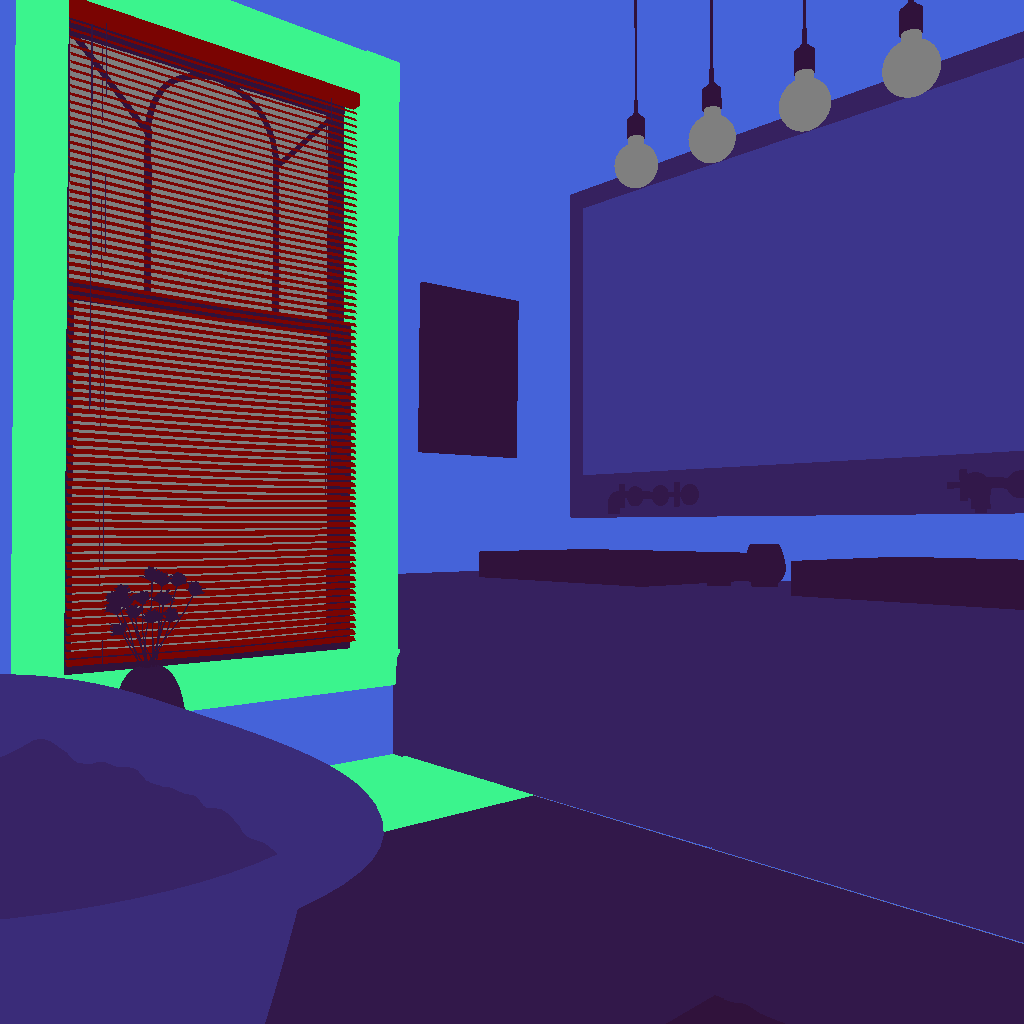}
  \caption{Objective camera}
  \label{fig:multiview-0}
\end{subfigure}
\hfill
\begin{subfigure}{0.32\columnwidth}
  \includegraphics[width=\linewidth]{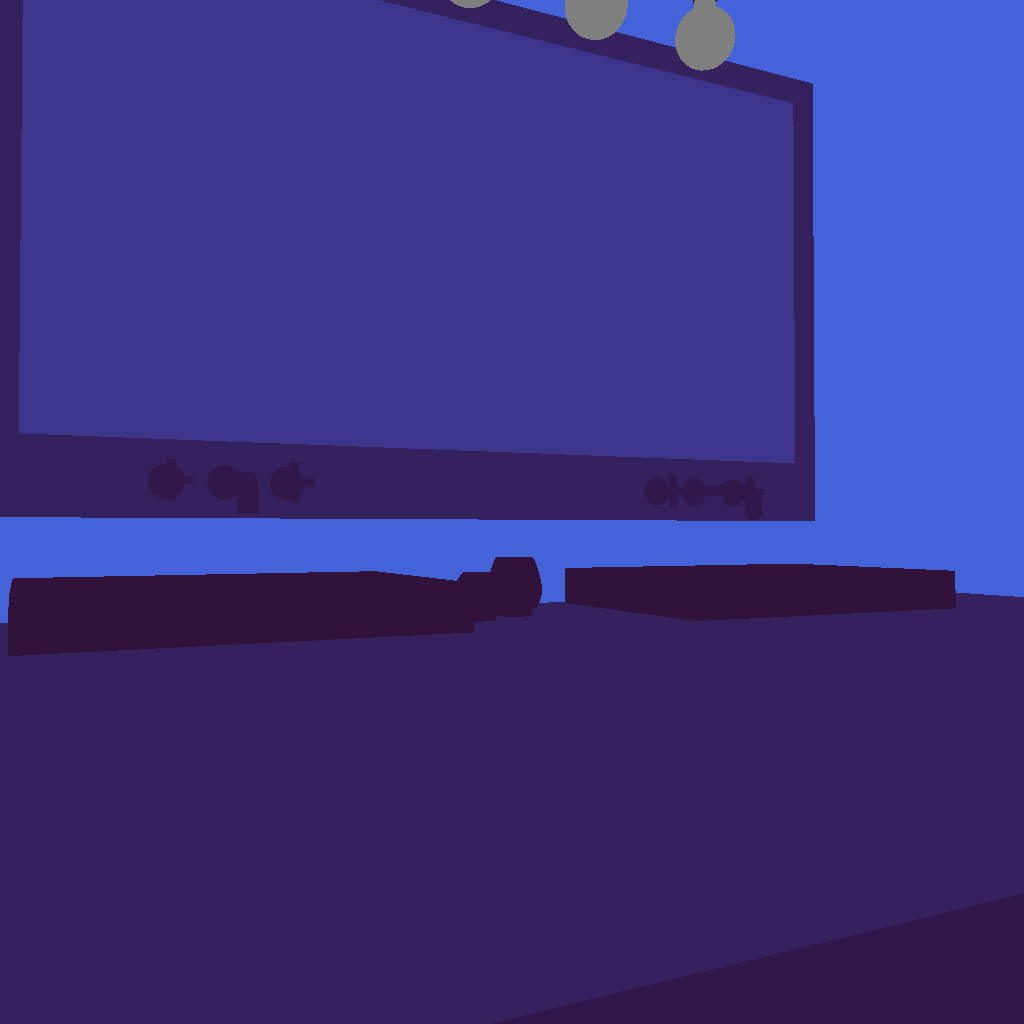}
  \caption{Inspection view 1}
  \label{fig:multiview-1}
\end{subfigure}
\hfill
\begin{subfigure}{0.32\columnwidth}
  \includegraphics[width=\linewidth]{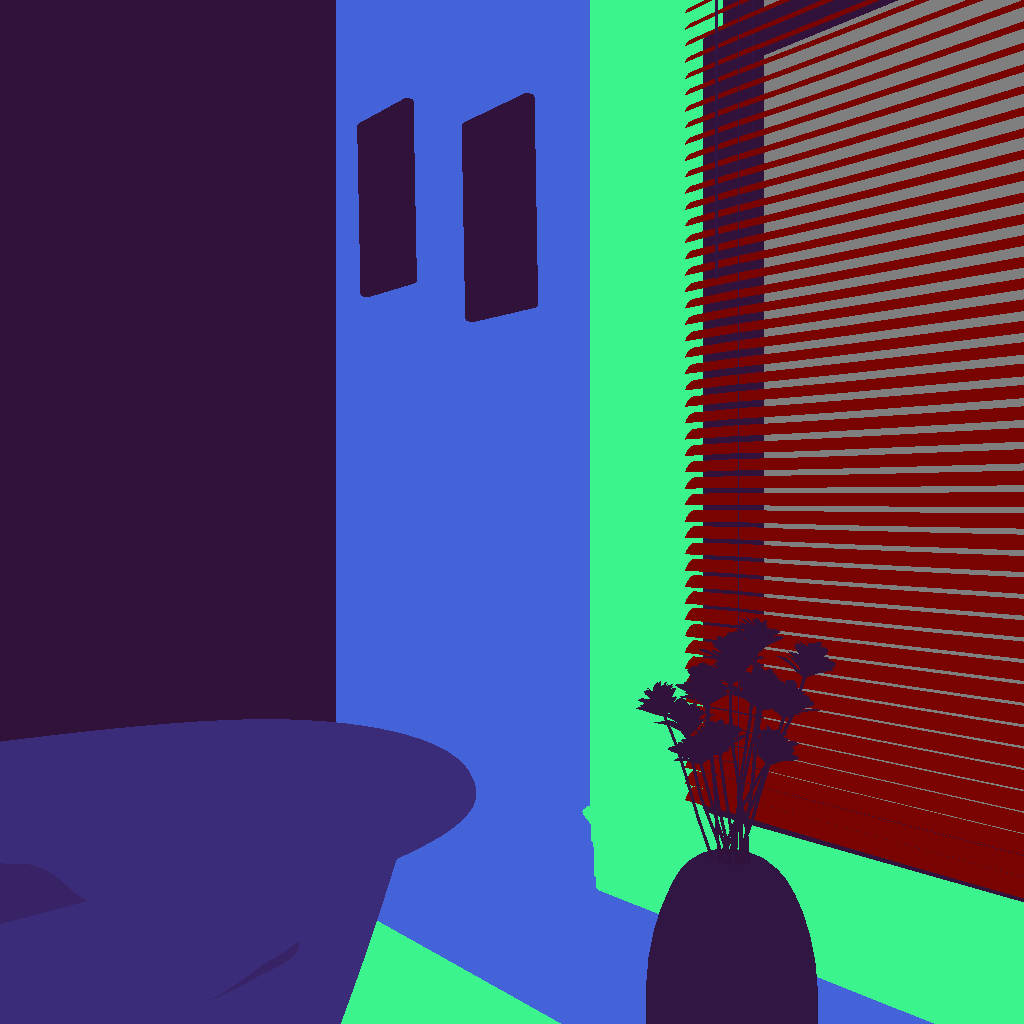}
  \caption{Inspection view 2}
  \label{fig:multiview-2}
\end{subfigure}
\caption{One buffer, many viewpoints. The buffer is computed once by an adjoint
pass fixed by the \emph{objective camera} (a). The
\emph{inspection camera} then moves freely
(b,~c): each view re-runs only
primary visibility and re-projects the \emph{same} sensitivities, with no
gradient recomputation. All three share one colour scale.}
\Description{The same object-level sensitivity buffer projected from three
different camera positions.}
\label{fig:multiview}
\end{figure}

Figure~\ref{fig:teaser} shows the result on an interior scene. Read at the
object level, the sensitivity AOV attributes the objective across the scene's
materials, including surfaces whose influence is carried by indirect transport
and is not evident in the primal image. Read as per-type channels it separates
which \emph{kind} of parameter drives the objective, and the hierarchy would
surface parameters such as a global exposure that never appear in the frame at
all. The forward channels of Figure~\ref{fig:duality} and the free-viewpoint
projection of Figure~\ref{fig:multiview} come from this same single pass.

\paragraph{Many objectives at once.} Because each objective conditions its own
sensitivity AOV, a scene can be inspected under several at once, one channel per
objective, so that a user selects among objective-conditioned derivative passes
exactly as they select among forward AOVs today. Objectives differ only in the
image-space adjoint $\partial M/\partial I$ that seeds the backward pass, but
they do not share that pass for free: each requires its own adjoint propagation
unless the renderer can carry several adjoint channels through one sweep, and
each populates its own buffer, so storage grows linearly in the number of
objectives inspected together.

\section{Implementation}
\label{sec:implementation}

We realize sensitivity AOVs on a physically based path tracer with reverse-mode
differentiation~\cite{Vicini2021PathReplay, Jakob2020DrJit}. Evaluating the
objective and backpropagating populates the buffer via
Equation~\ref{eq:chain}, a single pass that all placement views then share.
Binding a uniform parameter uses the object visible at each pixel, recovered
from a one-sample-per-pixel primary-visibility pass that also yields the UVs and
object identities the other views reuse. A spatially varying parameter carries
its per-texel gradient to the surface through these UVs; because the gradient
and the visible texel are sampled independently, we pool the per-texel field
before projection to fill gaps and suppress fireflies, which is the main
practical cost of the textured case. Hierarchy views are read directly from the
tree and need no image-space pass at all. The forward-mode view is the one
output not derived from the buffer: it requires its own forward-derivative
evaluation, seeding a single parameter and propagating to the image, and is
therefore recomputed per parameter shown. Everything else re-projects the one
reverse pass.

\section{Limitations and Future Work}
\label{sec:limitations}

The principal open view is the \emph{transport} view: rather than painting a
parameter where it lives, painting it where its influence \emph{lands}, so that
a wall's reflectance colors the table it brightens through interreflection.
Unlike the placement views, this cannot re-project a finished buffer: it must
accumulate contributions during the adjoint sweep and is tied to a single
viewpoint. We leave it, together with a fuller treatment of per-texel fields,
which is the least explored case in our current implementation, to future work.

Several further limitations bound the present method. Our placement views read
interior gradients and do not treat visibility discontinuities, so a
boundary-correct account of geometric sensitivity is out of scope here.
Sensitivities are local first-order quantities and do not capture interactions
between parameters. Like all Monte Carlo derivatives they inherit path-tracing
variance, so buffers computed at low sample counts may need denoising, and
denoising a sensitivity field for faithful inspection, without reordering which
parameters dominate, is itself an unstudied problem distinct from beauty-image
denoising. Finally, our objectives are deliberately simple; richer perceptual
and neural objectives fit the same framework and are natural to explore.

\section{Conclusion}
\label{sec:conclusion}

Differentiable renderers already compute, as a byproduct of every optimization,
a quantity that answers which parts of a scene matter for a given goal. We have
argued that this quantity deserves the same first-class output status as the
primal image. The sensitivity AOV names the output, the sensitivity buffer reduces each
further view to a primary-visibility pass over a stored result rather than a
second adjoint sweep, and
the selection, binding, and encoding choices determine how those derivatives are
shown to a human. Forward AOVs describe what the scene is; sensitivity AOVs
describe what would change it. Much of the space they open, the transport view,
per-texel fields, denoising for human inspection, remains unexplored, and we
offer this as scaffolding for it.

\begin{acks}
We acknowledge with gratitude the funding provided for this research by Simon
Fraser University as well as a Discovery Grant provided by the Natural Sciences
and Engineering Research Council of Canada. Additionally, we thank Benedikt
Bitterli~\cite{resources16} for providing the scene used in our figures.
\end{acks}

\bibliographystyle{ACM-Reference-Format}
\bibliography{references}

\end{document}